%% file: QuantumSAC.tex
\documentclass[conference]{IEEEtran}
\IEEEoverridecommandlockouts
\usepackage{amsmath,amssymb,amsfonts}
\usepackage{algorithm}
\usepackage{algorithmic}
\usepackage{graphicx}
\usepackage{textcomp}
\usepackage{xcolor}
\def\BibTeX{{\rm B\kern-.05em{\sc i\kern-.025em b}\kern-.08em
    T\kern-.1667em\lower.7ex\hbox{E}\kern-.125emX}}

\usepackage[round]{natbib}
\usepackage[T1]{fontenc}
\input{math_commands.tex}

\newcommand{\Actions}{\mathcal{A}}
\newcommand{\States}{\mathcal{S}}
\newcommand{\Prob}{p}
\usepackage{booktabs}
\usepackage{hyperref}
\usepackage{tikz}
\usetikzlibrary{quantikz}
\usepackage{subfigure}

\newcommand{\figwidthtwo}{0.48\textwidth}

\newcommand{\figwidthfour}{0.235\textwidth}

\begin{document}
\title{
  Variational Quantum Soft Actor-Critic\\
  \thanks{The source code is available at~\url{https://github.com/qlan3/QuantumExplorer}.}
}
\author{
  \IEEEauthorblockN{Qingfeng Lan}
  \IEEEauthorblockA{\textit{Department of Computing Science} \\
  \textit{University of Alberta}\\
  Edmonton, Canada \\
  qlan3@ualberta.ca}
}
\maketitle

\begin{abstract}
Quantum computing has a superior advantage in tackling specific problems, such as integer factorization and Simon's problem.
For more general tasks in machine learning, by applying variational quantum circuits, more and more quantum algorithms have been proposed recently, especially in supervised learning and unsupervised learning.
However, little work has been done in reinforcement learning, arguably more important and challenging.
Previous work in quantum reinforcement learning mainly focuses on discrete control tasks where the action space is discrete.
In this work, we develop a quantum reinforcement learning algorithm based on soft actor-critic --- one of the state-of-the-art methods for continuous control.
Specifically, we use a hybrid quantum-classical policy network consisting of a variational quantum circuit and a classical artificial neural network.
Tested in a standard reinforcement learning benchmark, we show that this quantum version of soft actor-critic is comparable with the original soft actor-critic, using much less adjustable parameters. 
Furthermore, we analyze the effect of different hyper-parameters and policy network architectures, pointing out the importance of architecture design for quantum reinforcement learning.\\
\end{abstract}

\begin{IEEEkeywords}
deep reinforcement learning, quantum reinforcement learning, quantum computation, variational quantum circuits
\end{IEEEkeywords}

\section{Introduction}

A classical bit can only represent $0$ or $1$.
However, a quantum bit or qubit can be in a superposition of both $0$ and $1$ simultaneously in quantum computing.
Furthermore, a multiple-qubit system can exhibit quantum entanglement, allowing a set of qubits to show a much more complex correlation than a classical system with the same number of bits.
These unique properties greatly enhance the power of quantum computation. Many quantum algorithms have been proposed to benefit from these properties, such as Grover's algorithm~\citep{grover1997quantum}, Simon's algorithm~\citep{simon1997power}, and Shor's algorithm~\citep{shor1994algorithms}.
These quantum algorithms are more efficient than classical algorithms for specific problems.

Recently, researchers in machine learning began to take advantage of quantum computing and invented many quantum machine learning algorithms, such as classification~\citep{farhi2020classification,schuld2020circuit,havlivcek2019supervised}, transfer learning~\citep{mari2020transfer}, generative adversarial networks~\citep{lloyd2018quantum,dallaire2018quantum}, and clustering problems~\citep{otterbach2017unsupervised}.
These algorithms utilize variational quantum circuits (VQCs), which are computation models with a sequence of quantum gates, wires, and measurements.
The quantum gates are usually ``variational'' with learnable parameters.
So VQCs are often viewed as the quantum version of artificial neural networks (ANNs), known as quantum neural networks or parameterized quantum circuits~\citep{beer2020training,broughton2020tensorflow}.

As another important area of machine learning, reinforcement learning (RL) has already shown its great success in Go~\citep{silver2017mastering}, video games~\citep{vinyals2019grandmaster,berner2019dota,zha2021douzero,badia2020agent57}, simulated robotic tasks~\citep{haarnoja2018soft,lillicrap2015continuous}, and recommendation systems~\citep{zheng2018drn,zhao2018deep,afsar2021reinforcement}.
Nevertheless, quantum RL remains to be an underexplored area. Most of the previous work focus on discrete control and these quantum algorithms are primarily based on Q-learing~\cite{williams1992simple}.
Moreover, not all the proposed quantum RL algorithms are properly tested in standard RL benchmarks. Instead, some of them are tested in sequential decision-making tasks transformed from classical problems, such as the eigenvalue problem and quantum state generation in quantum physics.

To fill the gaps mentioned previously, we propose variational quantum soft actor-critic --- a new quantum RL algorithm for continuous control tasks.
It is based on soft actor-critic (SAC)~\citep{haarnoja2018soft}, one of the state-of-the-art RL algorithms.
Specifically, we design a hybrid quantum-classical policy network consisting of a VQC and a classical ANN.
We test this algorithm in one of the standard RL tasks and show that it is able to achieve a similar performance compared to the classical SAC with much less learnable parameters.

\section{Background}

In this section, we first introduce some background knowledge of RL for continuous control. We then present a short introduction of soft actor-critic and a brief description of VQCs.

\begin{figure}[tbp]
\centering
\vspace{-0.2in}
\includegraphics[width=\figwidthtwo]{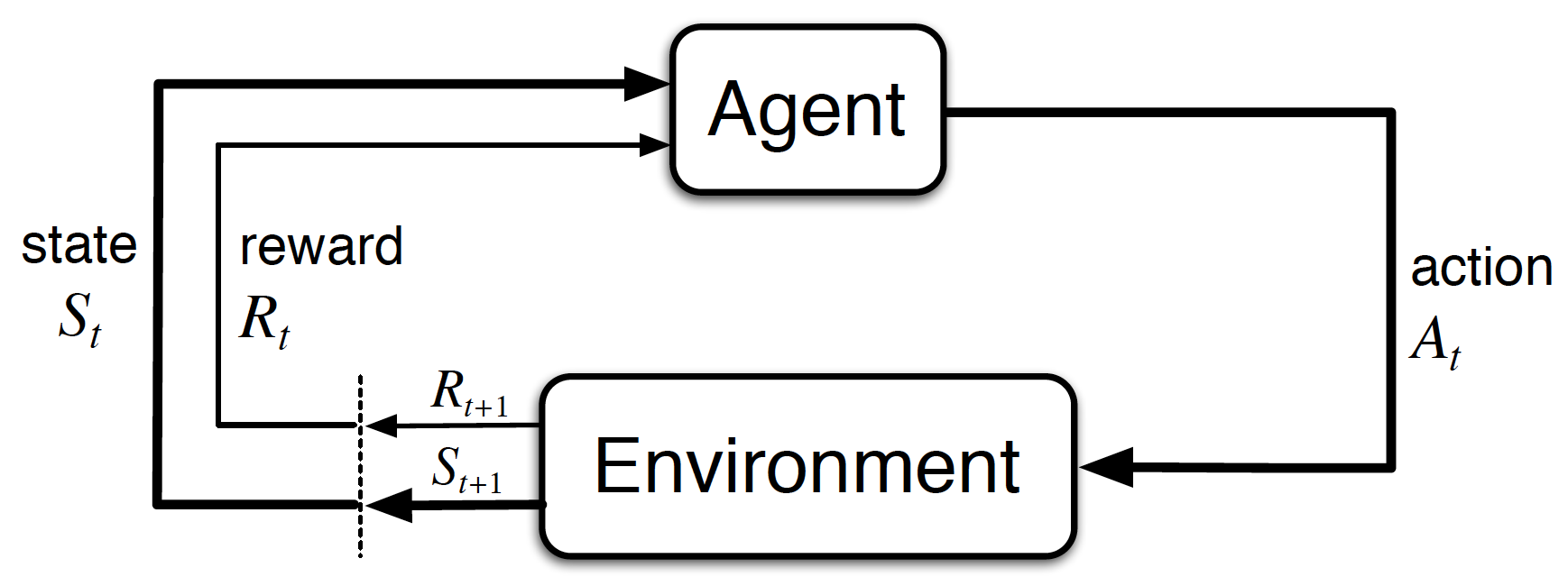}
\caption{The interaction circle between the agent and the environment in RL setting. This figure is taken from Chapter 3 of~\cite{sutton2011reinforcement}.}
\label{fig_rl}
\end{figure}

\subsection{Reinforcement Learning for Continuous Control}

Consider a Markov decision process (MDP), $M=(\States, \Actions, \Prob, \Prob_0, r, \gamma)$.
Both the state space $\States$ and the action space $\Actions$ are assumed to be continuous. 
$\Prob: \States \times \Actions \times \States \rightarrow [0, \infty)$ is the state transition probability.
$\Prob_0: \States \rightarrow [0, \infty)$ is the initial state probability.
$r: \States \times \Actions \rightarrow \real$ is the reward function.
Finally, $\gamma \in [0,1)$ is the discount factor.

Given a MDP $M$, an agent generates a trajectory based on a policy distribution $\pi: \States \times \Actions \rightarrow [0,\infty)$ by interacting with the environment.
Specifically, starting from a state $S_0 \sim \Prob_0(\cdot)$, the agent samples an action $A_t \in \Actions$ according to the policy $\pi$ (i.e. $A_t \sim \pi(\cdot|S_t)$) at each time-step $t = 0, 1, 2, \dots$, receives a reward signal $R_t = r(S_t, A_t)$, and observes the next state $S_{t+1}$ which is sampled from the transition function (i.e. $S_{t+1} \sim \Prob(\cdot | S_t, A_t)$). The overall interaction circle between the agent and the environment is shown in Fig.~\ref{fig_rl}.

For infinite horizon tasks, we can define return as the total discounted reward from time-step $t$: $G_t = \sum_{k=t}^{\infty} \gamma^{k-t} r(S_k, A_k)$.
Value functions can be defined as the expected return under policy $\pi$, $v_{\pi}(s) = \E_{\pi}[G_t | S_t=s]$; similarly, action-value functions are defined as $q_{\pi}(s, a) = \E_{\pi}[G_t | S_t=s, A_t=a]$.
The goal of the agent is to obtain a policy $\pi$ that maximizes the expected return starting from initial states.
Formally, let a policy $\pi_{\theta}$ be a differentiable function of a weight vector $\theta$. Our goal is to find $\theta$ that maximizes the objective funtion
\begin{align}
J(\theta) = \int p_0(s) v_{\pi_{\theta}}(s) \mathrm{d}s.
\end{align}

\subsection{Soft Actor-Critic}

SAC has shown its excellent performance and high sample efficiency in many continuous control tasks~\citep{haarnoja2018soft}.
As one of the algorithm based on the maximum entropy framework~\citep{ziebart2008maximum,toussaint2009robot,haarnoja2017reinforcement}, SAC aims to maximize the expected return and policy entropy simultaneously.
Different from the standard RL setting, for maximum entropy RL, the return includes entropy terms, defined as
\begin{align}
G_t = \sum_{k=t}^{\infty} \gamma^{k-t} (r(S_k, A_k) + \alpha \gH(\pi_{\theta}(\cdot|S_k)),
\end{align}
where $\alpha$ is a positive constant and $\gH(p) = - \int_{x} p(x) \log{p(x)} \de x$ is the differential entropy for probability density function $p(x)$.
The value functions and action-value functions are defined as $v_{\pi_{\theta}}(s_0) = \E_{\pi}\left[\sum_{t=0}^{\infty} \gamma^{t}(r(s_t, a_t) + \alpha \gH(\pi_{\theta}(\cdot | s_t)))\right]$ and $q_{\pi_{\theta}}(s_0, a_0) = \E_{\pi} \left[\sum_{t=0}^{\infty} \gamma^{t} r(s_t, a_t) + \alpha \sum_{t=1}^{\infty} \gamma^{t} \gH(\pi_{\theta}(\cdot | s_t)) \right]$.
The optimization objective is still defined as $J(\theta) = \int p_0(s) v_{\pi_{\theta}}(s) \mathrm{d}s$.
The policy entropy regularization is believed to improve exploration by encouraging more stochastic policies~\citep{haarnoja2017reinforcement,ziebart2010modeling}.
There is also evidence showing that the optimization landscape gets smoother by introducing entropy regularization, which allows a larger learning rate and makes it easier to optimize~\citep{ahmed2019understanding}.

In practice, we usually parameterize the action-value function with a neural network as $Q_{\phi}(s, a)$, where $\phi$ is the weight vector of the neural network. As shown in \cite{haarnoja2018soft}, the objective can then be further reduced to
\begin{align}
J(\theta) = \E_{S \sim d_{\pi_{\theta}}, A \sim \pi_{\theta}}[Q_{\phi}(S,A) - \alpha \log(\pi_{\theta}(A|S))],
\end{align}
where $d_{\pi_{\theta}}$ is the state distribution given policy $\pi_{\theta}$.
Furthermore, to reduce the overestimation bias, two action-value functions ($Q_{\phi_1}$ and $Q_{\phi_2}$) are introduced and the minimum action-value is used~\citep{fujimoto2018addressing}. The reparameterization technique is applied to reduce the gradient estimation variance. 
To be specific, the action $A$ is sampled from the policy $\pi$ parameterized with $\theta$ given the current state $S$: $A \sim \pi_{\theta}(\cdot | S)$. 
We reparameterize the action with a function $f$, $A = f_{\theta}(\eps;S), \, \eps \sim p(\cdot)$ where $p(\cdot)$ is some fixed distribution such as Gaussian.
For simplicity, let $\tilde{A}_{\theta}=f_{\theta}(\eps;S)$ when $\eps$ and $S$ can be easily deduced given the context. Moreover, since $d_{\pi_{\theta}}$ is hard to compute in practice, we use a replay buffer $\gD$ to approximate it. Combining the above modifications, we then rewrite the objective as:
\begin{align}
J(\theta) = \underset{S \sim \gD, \eps \sim \gN(0,1)}{\E}[\min_{i=1,2}Q_{\phi_i}(S,\tilde{A}_{\theta}) - \alpha \log(\pi_{\theta}(\tilde{A}_{\theta}|S))].
\end{align}

\begin{figure*}[tbp]
\vspace{-0.2in}
\centerline{
\begin{quantikz}
& \gate{R_x(s_1)} \gategroup[3,steps=1,style={dashed,rounded corners,fill=cyan, inner xsep=2pt},background]{encoder layer} & \gate{R(\alpha_1^1, \beta_1^1, \gamma_1^1)} \gategroup[3,steps=4,style={dashed,rounded corners,fill=green!50, inner xsep=2pt},background]{1st variational layer} & \ctrl{1} & \qw      & \targ{}   & \gate{R(\alpha_1^2, \beta_1^2, \gamma_1^2)} \gategroup[3,steps=4,style={dashed,rounded corners,fill=green!50, inner xsep=2pt},background]{2nd variational layer} & \ctrl{2} & \targ{}  & \qw   & \meter{} & \qw \\
& \gate{R_x(s_2)} & \gate{R(\alpha_2^1, \beta_2^1, \gamma_2^1)} & \targ{}  & \ctrl{1} & \qw       & \gate{R(\alpha_2^2, \beta_2^2, \gamma_2^2)} & \qw      & \ctrl{-1} & \targ{} & \meter{} & \qw \\
& \gate{R_x(s_3)} & \gate{R(\alpha_3^1, \beta_3^1, \gamma_3^1)} &  \qw     & \targ{}  & \ctrl{-2} & \gate{R(\alpha_3^2, \beta_3^2, \gamma_3^2)} &  \targ{}  & \qw  & \ctrl{-1} & \meter{} & \qw
\end{quantikz}
}
\caption{The architecture of a vanilla VQC with 3 qubits.}
\label{fig_VanillaVQC}
\end{figure*}
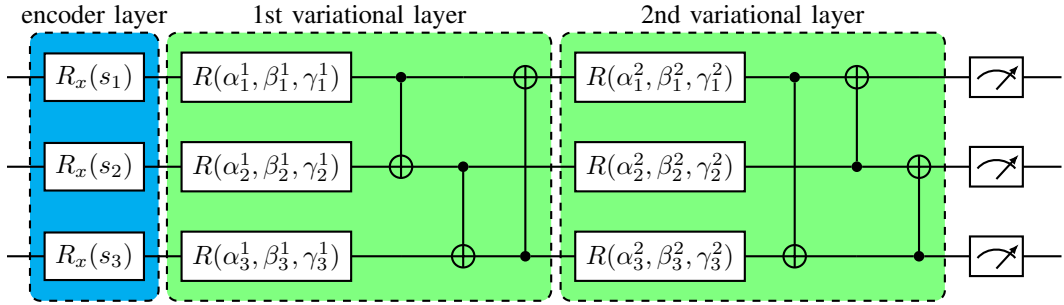

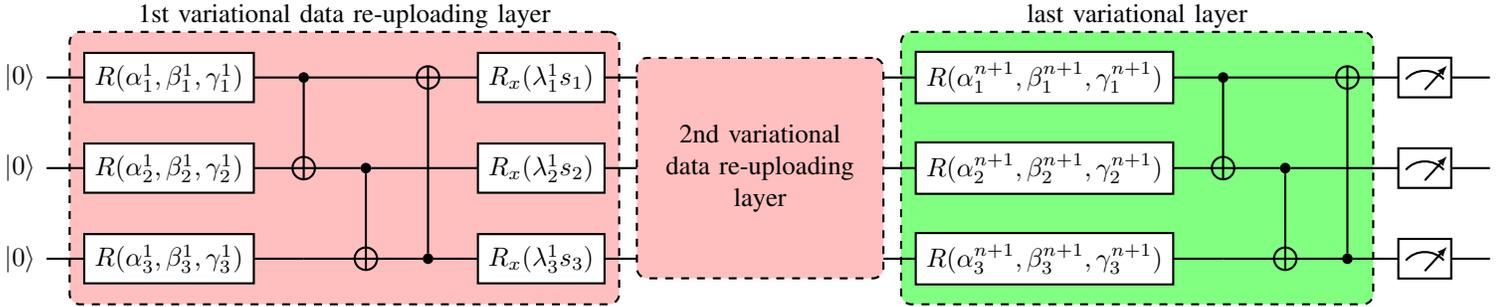
\begin{figure*}[tbp]
\centerline{
\begin{quantikz} 
& \lstick{\ket{0}} & \gate{R(\alpha_1^1, \beta_1^1, \gamma_1^1)} \gategroup[3,steps=5,style={dashed,rounded corners,fill=pink, inner xsep=2pt},background]{1st variational data re-uploading layer} & \ctrl{1} & \qw      & \targ{}   & \gate{R_x(\lambda_1^1 s_1)} & \gate[wires=3,style={dashed,rounded corners,fill=pink, inner xsep=2pt},background,disable auto height]{\begin{array}{c}\text{2nd variational} \\ \text{data re-uploading} \\ \text{layer} \end{array}} & \gate{R(\alpha_1^{n+1}, \beta_1^{n+1}, \gamma_1^{n+1})} \gategroup[3,steps=4,style={dashed,rounded corners,fill=green!50, inner xsep=2pt},background]{last variational layer} & \ctrl{1} & \qw  & \targ{}     & \meter{} & \qw \\
& \lstick{\ket{0}} & \gate{R(\alpha_2^1, \beta_2^1, \gamma_2^1)} & \targ{}  & \ctrl{1} & \qw       & \gate{R_x(\lambda_2^1 s_2)} & & \gate{R(\alpha_2^{n+1}, \beta_2^{n+1}, \gamma_2^{n+1})} & \targ{}  & \ctrl{1} & \qw & \meter{} & \qw \\
& \lstick{\ket{0}} & \gate{R(\alpha_3^1, \beta_3^1, \gamma_3^1)} &  \qw     & \targ{}  & \ctrl{-2} & \gate{R_x(\lambda_3^1 s_3)} & & \gate{R(\alpha_3^{n+1}, \beta_3^{n+1}, \gamma_3^{n+1})} &  \qw  & \targ{}  & \ctrl{-2} & \meter{} & \qw
\end{quantikz}
}
\caption{The architecture of a data re-uploading VQC with 3 qubits.}
\label{fig_ReUploadingVQC}
\end{figure*}

\subsection{Variational Quantum Circuits}

A VQC consists of various quantum gates and wires, with learnable parameters~\citep{benedetti2019parameterized} to adjust these gates (e.g. rotation gates).
Similar to parameters in ANNs, these learnable parameters in VQCs can be optimized to approximate complex continuous functions.
So VQCs are also known as quantum neural networks or parameterized quantum circuits~\citep{beer2020training,broughton2020tensorflow}. 

A typical VQC mainly has three components --- an encoder circuit $U_{\theta_{enc}}$, a variational circuit $U_{\theta_{var}}$, and measurement.
The encoder circuit is usually the first operation of a quantum circuit.
Given a classical input $\emph{x}$, $U_{\theta_{enc}}$ encodes it into a quantum state: $\emph{x} \rightarrow U_{\theta_{enc}}(\emph{x}) |0\rangle^{\bigotimes n}$, where $n$ is the number of qubits.
Here, we introduce three encoder circuits --- basis embedding, angle embedding, and amplitude embedding.
The basis embedding encodes $n$ binary features into a basis state of $n$ qubits.
The angle embedding encodes $n$ features into rotation angles of $n$ qubits with the help of rotation gates.
Finally, the amplitude embedding encodes $2^n$ features into the amplitude vector of $n$ qubits, which provides an exponential reduction in terms of the number of qubits~\citep{benedetti2019parameterized}.
A variational circuit often follows after an encoder circuit. Usually, it consists of alternate layers of rotation gates and entanglement gates (e.g., a closed chain or ring of CNOT gates)~\citep{kandala2017hardware,schuld2020circuit}. After a variational circuit, we measure the states of qubits in the end.

Similar to an ANN, we can also train a VQC with gradient descent. Given a quantum simulator, we could apply backpropagation to compute gradient analytically and optimize the VQC parameters. However, backpropagation is not applicable for a physical quantum computer since it is impossible to measure and store intermediate quantum states during computation without impacting the whole computation process. Instead, we apply (stochastic) parameter-shift~\citep{li2017hybrid,mitarai2018quantum,schuld2019evaluating,banchi2021measuring} to compute gradients for any multi-qubit VQCs. After getting gradients, we can then optimize the parameters with traditional optimizers used for ANNs, such as RMSprop~\citep{tieleman2012} and Adam~\citep{kingma2015adam}.

\section{Related Work}

There are already many applications of VQCs and hybrid quantum-classical systems in quantum machine learning~\citep{biamonte2017quantum}, such as supervised learning~\citep{rebentrost2014quantum,amin2018quantum}, unsupervised learning~\citep{basheer2020quantum,winci2020path}, and RL~\citep{dunjko2017advances,kwak2021introduction}. Since this work is mainly about applying a VQC to RL, we will mainly introduce some related work in quantum RL in this section.

\cite{dong2008quantum} suggested to represent states and actions as eigenvectors of a Hilbert space for a quantum system and proposed a quantum RL algorithm based on Q-learning~\citep{dayan1992q}. This algorithm applies Grover iteration~\citep{grover1997quantum} to update action selection probabilities. It is shown to have an excellent exploration-exploitation balance in a gridworld environment. However, it is only applicable to tasks with discrete states and actions, due to the discrete representation scheme.
\cite{wang2020deep} designed a quantum version of the classical cartpole balancing problem~\citep{barto1983neuronlike} and showed that deep Q-learning~\citep{mnih2013playing,mnih2015human} was able to solve this task.
The results suggested that general quantum control tasks can also serve as RL benchmarks.

\cite{chen2020variational} and \cite{skolik2021quantum} showed that variational quantum deep Q-learning worked with the experience replay and the target network. 
\cite{kwak2021introduction} proposed a quantum RL algorithm based on deep Q-learning~\citep{mnih2015human} and proximal policy optimization~\citep{schulman2017proximal}.
The implementation used the PennyLane library~\citep{bergholm2018pennylane} and was tested in the cartpole environment~\citep{barto1983neuronlike}.
Later, softmax-PQC is developed by \cite{jerbi2021parametrized} which applied a softmax policy to successfully solve several standard RL tasks, based on REINFORCE~\citep{williams1992simple}.
\cite{wu2020quantum} extended quantum RL algorithm to tasks with continuous action spaces based on deep deterministic policy gradient (DDPG)~\citep{silver2014deterministic}.
Finally, \cite{wei2021deep} proposed a deep RL algorithm with a quantum-inspired experience replay which helps to achieve a better balance between exploration and exploitation.

\begin{figure}[tbp]
\centering
\includegraphics[width=\figwidthtwo]{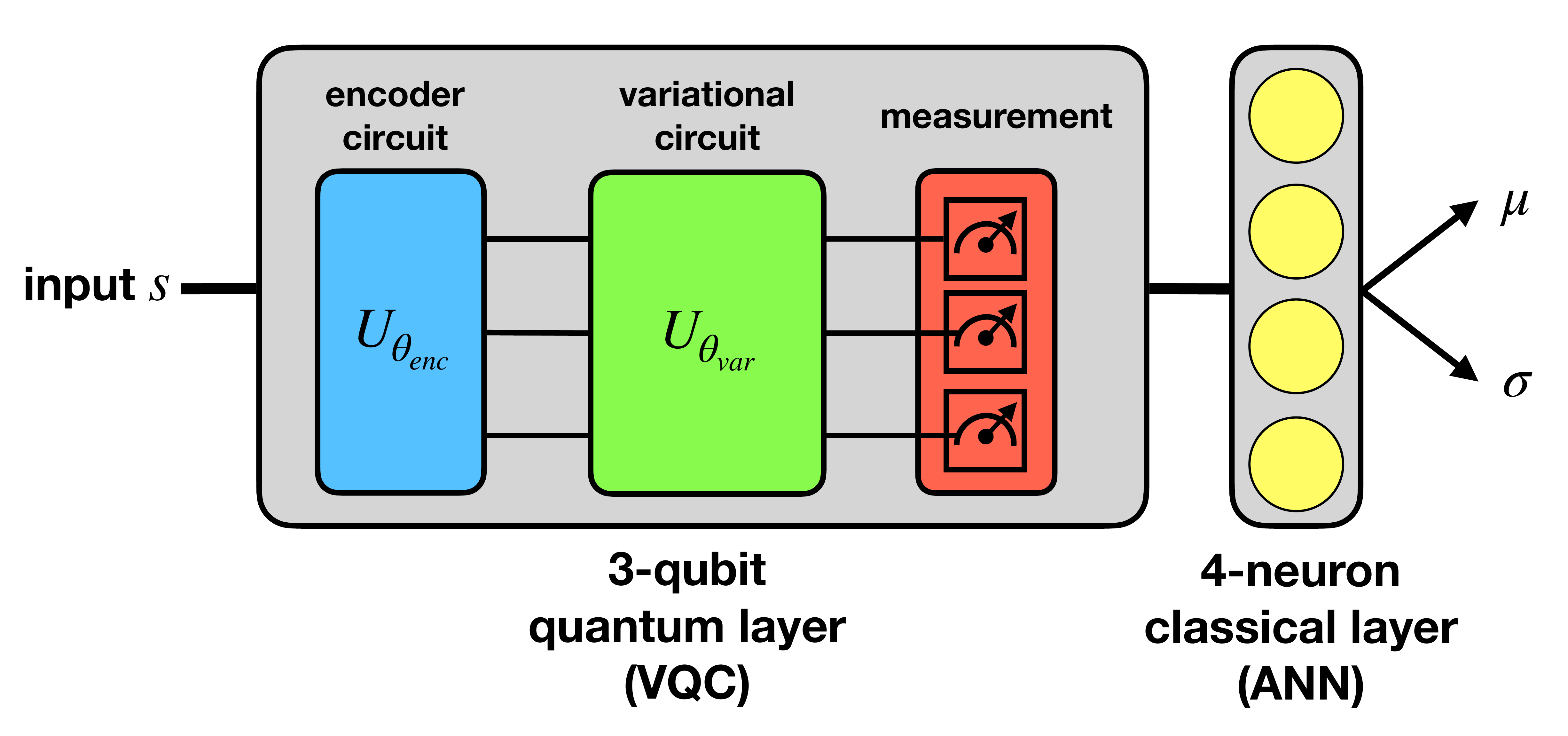}
\caption{The architecture of a hybrid quantum-classical policy network for QuantumSAC.}
\label{fig_HybirdNN}
\end{figure}

\section{Variational Quantum Soft Actor-Critic}

Although \cite{wu2020quantum} proposed the quantum DDPG algorithm to solve continuous tasks, it is not tested in any standard RL benchmarks.
To fill this gap, we propose variational quantum SAC (QuantumSAC) --- a quantum version of SAC with a hybrid quantum-classical policy network.
Note that for both SAC and QuantumSAC, a policy network has state $s$ as input and outputs the parameters of a policy distribution.
For example, we output the mean $\mu$ and the standard deviation $\sigma$ for a Gaussian policy distribution.
For QuantumSAC, following~\cite{jerbi2021parametrized}, we use a VQC to encode input state $s$ by angle embedding where each qubit is used to encode one element in $s$.
Thus, the number of qubits equals the dimension of state $s$.
Since the input ($s$) dimension and the output ($\mu$ and $\sigma$) dimension are usually different, we may encounter a dimension mismatch if we measure the state of each qubit and output the results directly.
To solve this problem, after measuring the states of qubits, we apply a linear ANN to transform the measurement results to parameters of a policy distribution.
The overall structure of an examplar hybrid quantum-classical neural network (with three qubits and four neurons) is shown in Fig.~\ref{fig_HybirdNN}.

We present a complete description of QuantumSAC in Algorithm~\ref{algo_qsac}. The critical difference between SAC and QuantumSAC is that we replace the classical policy network in SAC with a hybrid quantum-classical policy network in QuantumSAC.
In general, hybrid quantum-classical networks can be used to estimate action-value functions as well. However, using a hybrid quantum-classical policy network is enough to serve our purpose, and we still use ANNs as action-value networks for simplicity.

\begin{algorithm}[tbp]
\caption{QuantumSAC} \label{algo_qsac}
\begin{algorithmic}
  \STATE \textbf{Input:} initial policy parameters $\theta$, initial action-value estimate parameters $\phi_1$ and $\phi_2$, $\gamma$, $\alpha$, $\rho$, empty experience replay $\gD$.
  \STATE Initialize the hybrid quantum-classical policy network with $\theta$. 
  \STATE Initialize two action-value networks with $\phi_1$ and $\phi_2$ respectively.
  \STATE Set target action-value networks parameters: $\phi_{targ,1} \leftarrow \phi_1$ and $\phi_{targ,2} \leftarrow \phi_2$.
  \FOR{each time-step}
    \STATE Observe state $S$, select action $A \sim \pi_{\theta}(\cdot|S)$, and excute $A$ in the environment.
    \STATE Observe next state $S'$, reward $R$, and binary done signal $d$ to indicate whether $S'$ is a terminal state or not.
    \STATE Store $(S,A,R,S',d)$ in $\gD$.
    \STATE Reset the environment state if $d=1$ (i.e. $S'$ is terminal).
    % Update parameters
    \STATE Sample a batch of transitions $B=\{(S,A,R,S',d)\}$ from $\gD$ randomly.
    \STATE Compute target values $y(R,S',d) = R + \gamma (1-d) \left(\min_{i=1,2} Q_{\phi_{targ,i}}(S', A') - \alpha \log{\pi_{\theta}(A'|S')} \right)$ where $A' \sim \pi_{\theta}(\cdot|S')$.
    \STATE Update $\phi_i$ by minimizing: $\E_{B}[(Q_{\phi_i}(S,A) - y(R,S',d))^2]$ for $i=1,2$.
    \STATE Update $\theta$ by maximizing: $\E_{B}[\min_{i=1,2} Q_{\phi_i}(S,\tilde{A}_{\theta}) - \alpha \log{\pi_{\theta}(\tilde{A}_{\theta}|S)}]$.
    \STATE Do a soft update for target action-value networks: $\phi_{targ,i} \leftarrow \rho \phi_{targ,i} + (1-\rho) \phi_i$ for $i=1,2$.
  \ENDFOR
\end{algorithmic}
\end{algorithm}

\begin{figure}[tbp]
\centering
\vspace{-0.2in}
\includegraphics[width=\figwidthfour]{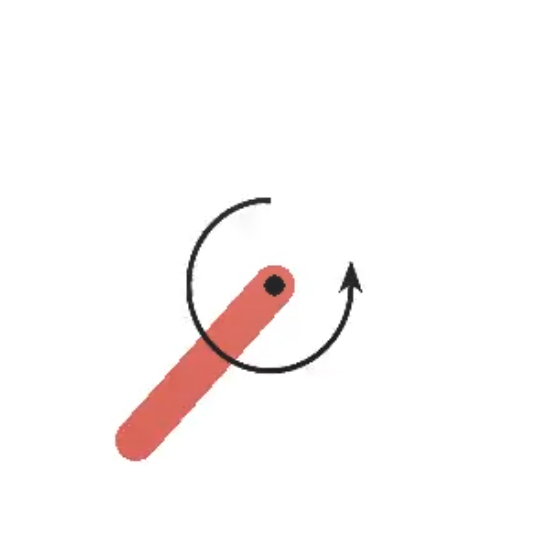}
\vspace{-0.2in}
\caption{Pendulum-v0 from OpenAI gym. The goal of this task is to swing a frictionless pendulum up and keep it upright. The state is a $3$d continuous vector, and the action is a real number chosen from $[-2.0,2.0]$. For each episode, the pendulum starts at a random angle from $-\pi$ to $\pi$ with a random velocity between $-1$ and $1$. After $200$ steps, an episode terminates.}
\label{fig_pendulum}
\end{figure}

\section{Experiments}

In this section, we demonstrate the potential of QuantumSAC in a standard RL benchmark from OpenAI gym~\citep{brockman2016openai} --- Pendulum-v0 (Fig.~\ref{fig_pendulum}).
Both the state space and action space for Pendulum-v0 are continuous.
The state is a $3$d vector, and the action is a real number chosen from $[-2.0,2.0]$. 
So we use a linear layer with input dimension $3$ and output dimension $2$ ($\mu$ and $\sigma$) for the ANN part in the hybrid quantum-classical policy network.
For each episode, the pendulum starts at a random angle from $-\pi$ to $\pi$ with a random velocity between $-1$ and $1$. After $200$ steps, an episode terminates.
The goal of this task is to swing a frictionless pendulum up and keep it upright.

For the VQC part, we investigate two options, a vanilla VQC and a data re-uploading VQC~\citep{perez2020data}, as shown in Fig.~\ref{fig_VanillaVQC} and Fig.~\ref{fig_ReUploadingVQC} respectively. 
A vanilla VQC has an encoder circuit (an angle embedding layer) and several variational layers without any other fancy structures. A data re-uploading VQC consists of several variational data re-uploading layers and one last variational layer. Compared to a vanilla VQC, the input data is encoded many times in a data re-uploading VQC, which significantly improves the expressivity of VQCs~\citep{perez2020data,skolik2021quantum,jerbi2021parametrized}. We will verify this claim in our settings.

The experimental setup is as follows.
Our implementations of SAC and QuantumSAC were based on PyTorch~\citep{paszke2019pytorch} and PennyLane~\citep{bergholm2018pennylane}, running in simulation.
We trained each algorithm with $50,000$ steps.
For both algorithms, the action-value networks were multi-layer perceptrons (MLPs) with hidden layers $[32,32]$.
The policy network for SAC was also an MLP with hidden layers $[32,32]$.
For QuantumSAC, the policy network was a hybrid quantum-classical network where the number of VQC layers~\footnote{For a vanilla VQC, the number of VQC layers refers to the number of variational layers. For a data re-uploading VQC, the number of VQC layers refers to the number of variational data re-uploading layers.} (denoted as $n$) was chosen from $\{1,2,4,8\}$.
The size of the experience replay was $10,000$. The batch size was $32$.
The discount factor $\gamma$ was $0.99$. $\alpha$ was set to $0.2$ and $\rho$ was set to $0.995$.
All networks were optimized by Adam~\citep{kingma2015adam}.
The step-size of action-value networks were set to $3e-3$. 
We did a grid search for the step-size of policy networks with scope $\{1e-1, 3e-2, 1e-2, 3e-3, 1e-3, 3e-4\}$.
We compared the performance of each algorithm by the mean return of the last $10$ training episodes.
The depicted returns in all learning curves were averaged over the last $10$ episodes, and the curves were smoothed using an exponential average. All experimental results were averaged over $10$ runs, with the shaded area representing one standard error.

\begin{figure}[tbp]
\centering
\vspace{-0.2in}
\includegraphics[width=\figwidthtwo]{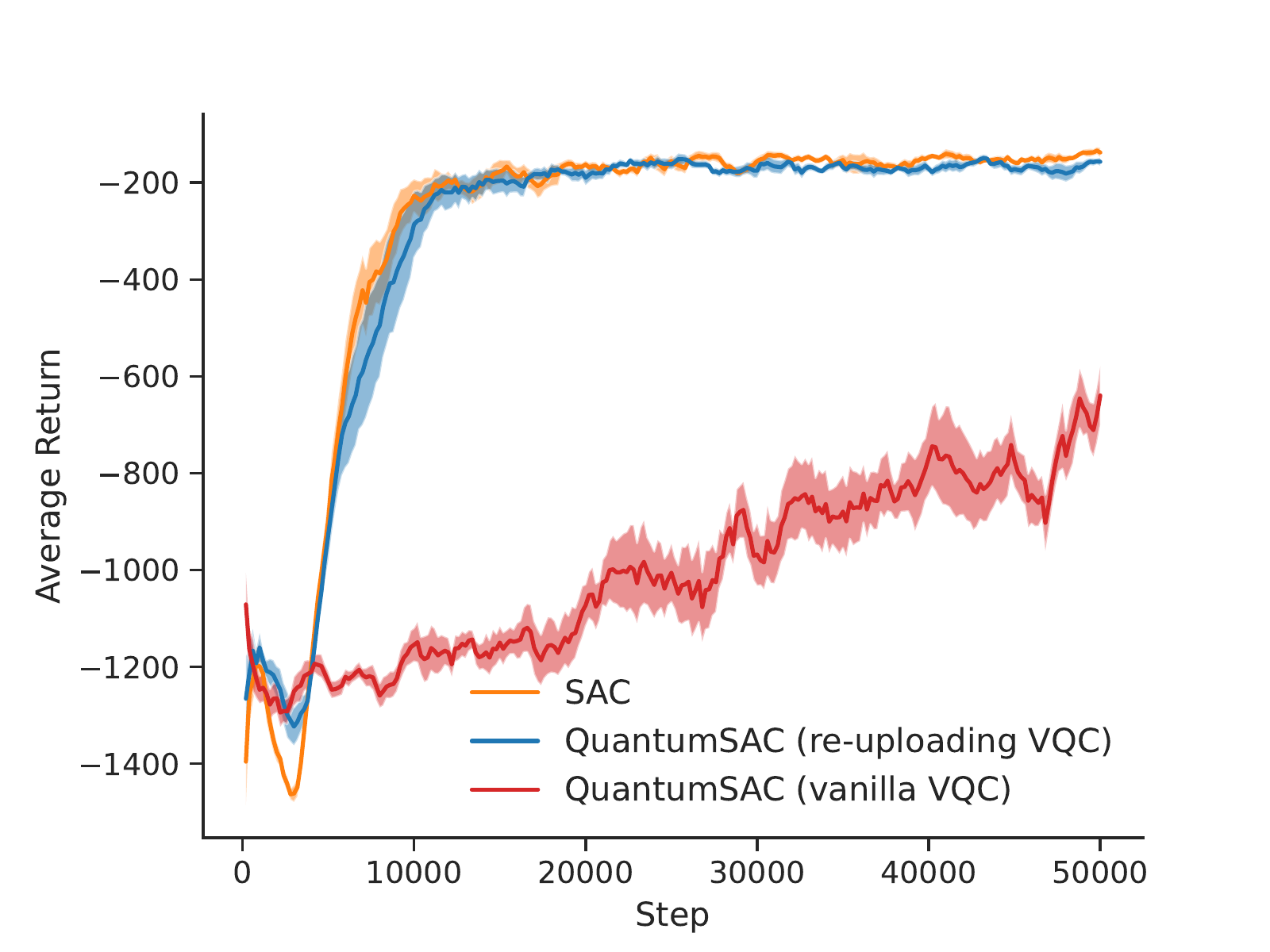}
\caption{Learning curves of QuantumSAC and SAC with the best hyper-parameter setting in Pendulum-v0. The depicted return in all learning curves was averaged over the last $10$ episodes, and the curves were smoothed using an exponential average. All experimental results were averaged over $10$ runs, with the shaded area representing one standard error. In terms of the overall performance, SAC $\approx$ QuantumSAC (re-uploading VQC) $>$ QuantumSAC (vanilla VQC).}
\label{fig_best}
\end{figure}

\paragraph{QuantumSAC v.s. SAC}

We first show learning curves of QuantumSAC and SAC with their best hyper-parameter settings in Fig.~\ref{fig_best}.
Clearly, we see that with a data re-uploading VQC policy network, QuantumSAC almost performs as well as SAC.
Note that the number of VQC layers $n$ of QuantumSAC (re-uploading VQC) shown in this figure is $2$.
So the total number of learnable parameters in this data re-uploading VQC policy network is just $41$, while the total number of learnable parameters in SAC policy network is $1,250$! This result shows the significant advantage of VQCs in reducing model parameters while achieving a comparable sample efficiency.

\begin{figure}[tbp]
\vspace{-0.2in}
\begin{center}
\subfigure[QuantumSAC with a vanilla VQC]{
  \includegraphics[width=\figwidthtwo]{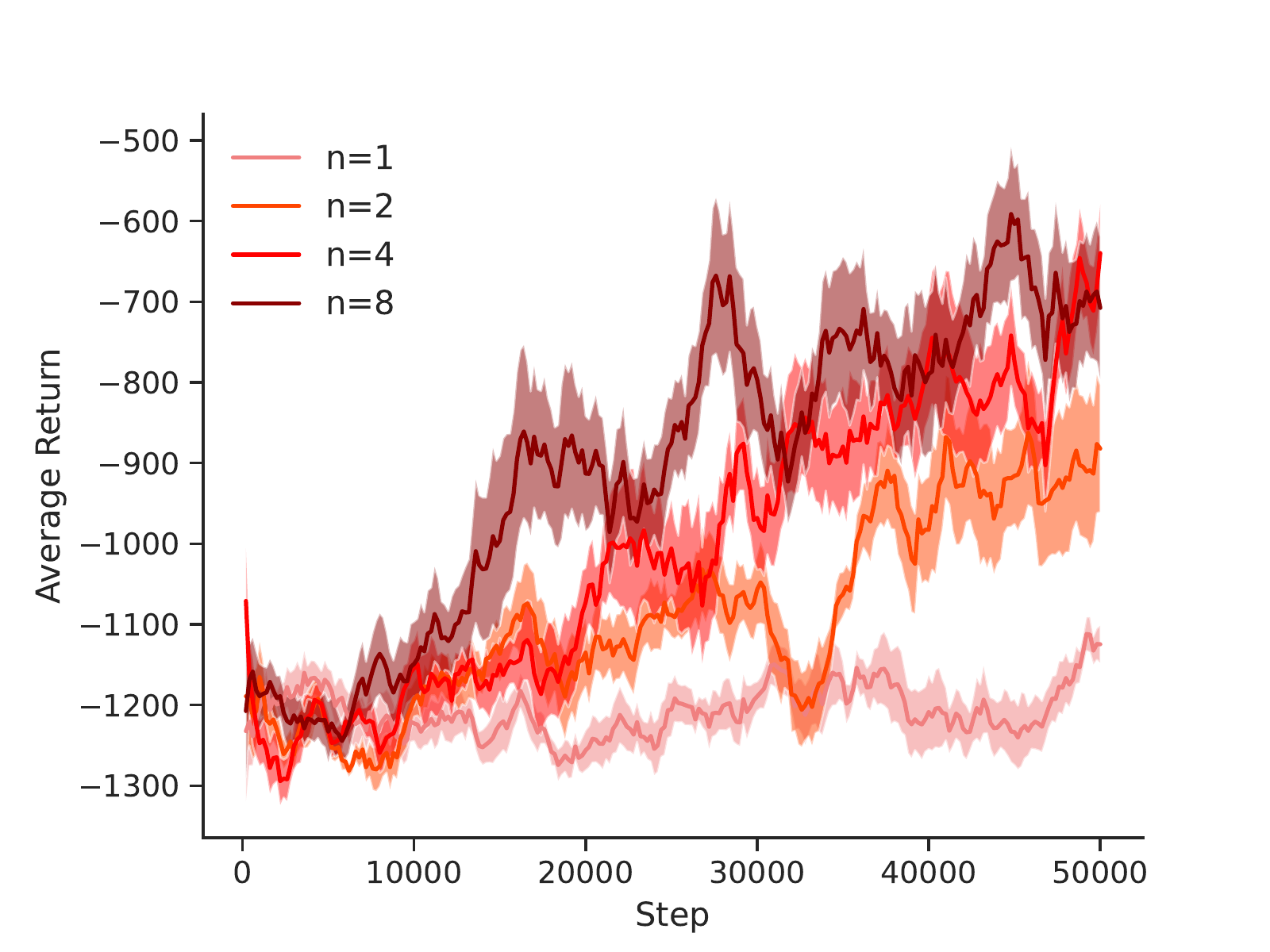}
}
\subfigure[QuantumSAC with a data re-uploading VQC]{
  \includegraphics[width=\figwidthtwo]{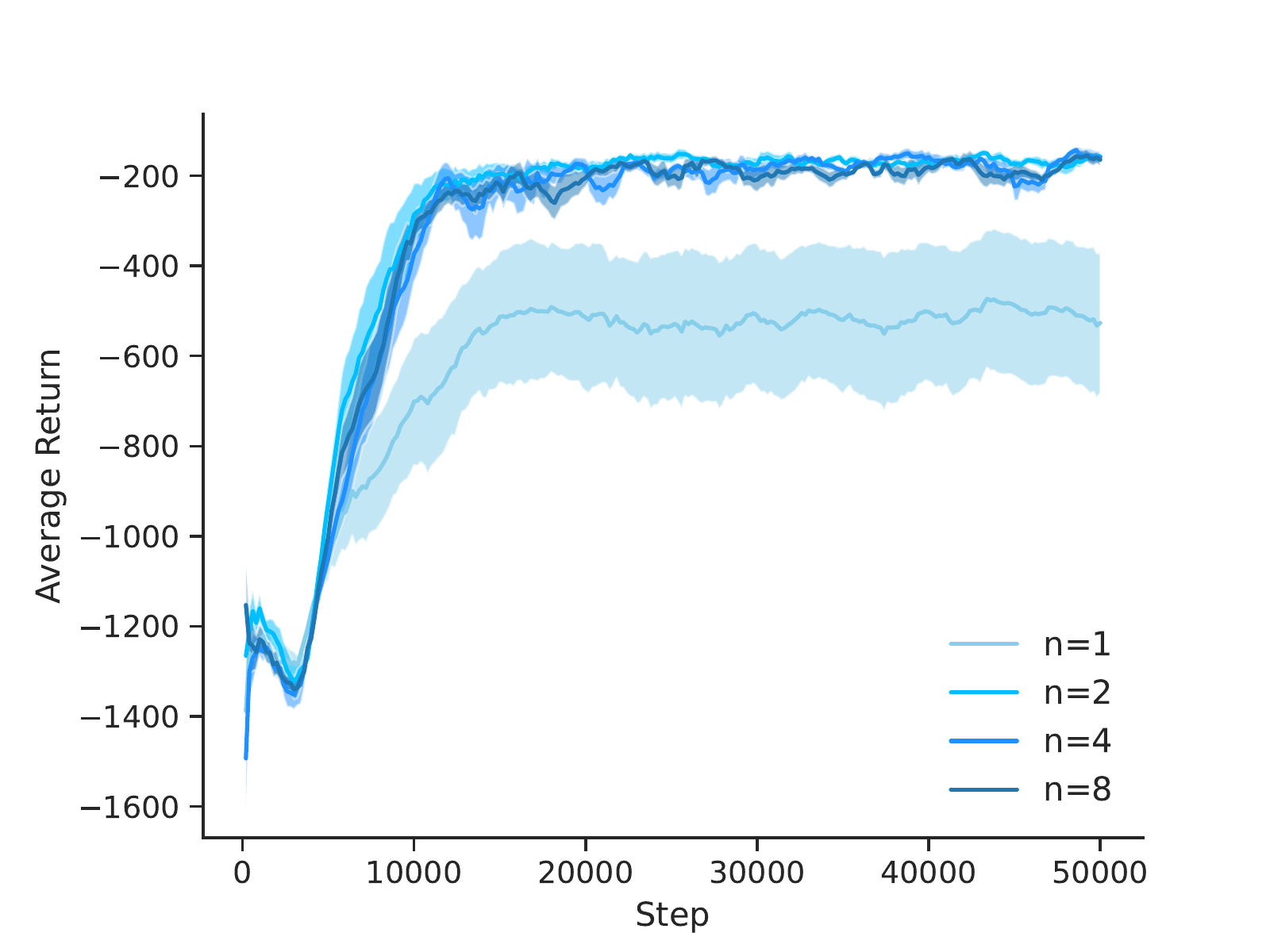}
}
\end{center}
\vspace{-0.1in}
\caption{Learning curves of QuantumSAC with various numbers of VQC layers in Pendulum-v0. The depicted return in all learning curves was averaged over the last $10$ episodes, and the curves were smoothed using an exponential average. All experimental results were averaged over $10$ runs, with the shaded area representing one standard error. In general, more VQC layers lead to better performance.}
\label{fig_layers}
\end{figure}

However, the performance of QuantumSAC is greatly influenced by the architecture of the VQC --- QuantumSAC (vanilla VQC) is outperformed by both SAC and QuantumSAC (re-uploading VQC). This result reveals the importance of architecture design for VQCs in quantum RL. The design of VQCs may even benefit from the evolution road of ANNs. To conclude, we believe that more empirical effort and theoretical understanding are needed in designing more powerful and efficient VQCs for quantum RL.

\paragraph{The influence of the number of VQC layers}

We varied the number of VQC layers $n$ in QuantumSAC for both vanilla VQCs and data re-uploading VQCs to study its effect on the learning process. To be specific, $n$ was chosen from $\{1,2,4,8\}$. 
For each $n$, we swept the step-size for the policy network and presented the results with the best hyper-parameter configurations. 
As shown in Fig.~\ref{fig_layers}, the performance is generally improved as $n$ increases. 
However, the detailed effect is different between vanilla VQCs and data re-uploading VQCs. 
For vanilla VQCs, there is a consistent performance improvement as $n$ increases, although the learning stability has no apparent change. 
For data re-uploading VQCs, there is a clear performance improvement when $n$ is increased from $1$ to $2$. 
The variance of the average return is also reduced obviously. 
However, after adding more layers, both the performance improvement and variance reduction are not evident. 
This result suggests that a $2$-layer data re-uploading VQC already has enough capacity to model the optimal policy, while the modelling capacity of an 8-layer vanilla VQC is much lower.

\section{Conclusion and Future Work}

In this work, we presented the variational quantum soft actor-critic algorithm and tested it in a pendulum balancing task.
We showed the quantum advantage in reducing model parameters while achieving similar performance to the classical algorithm.
To fully exploit the power of quantum computation in RL, designing more expressive and efficient VQCs is essential. As for future work, we would like to test our algorithm further in a physical quantum computer.

% \newpage
\bibliography{reference}
\end{document}

%% file: math_commands.tex
\usepackage{amsmath,amsfonts,bm}

\def\eqref#1{equation~\ref{#1}}
\def\1{\bm{1}}

\def\eps{{\epsilon}}

\DeclareMathAlphabet{\mathsfit}{\encodingdefault}{\sfdefault}{m}{sl}
\SetMathAlphabet{\mathsfit}{bold}{\encodingdefault}{\sfdefault}{bx}{n}

\def\gD{{\mathcal{D}}}

\def\gH{{\mathcal{H}}}

\def\gN{{\mathcal{N}}}

\newcommand{\real}{\mathbb{R}}
\newcommand{\E}{\mathbb{E}}

\newcommand{\de}{\mathrm{d}}